# Nonlinear optical performances of supramolecular 1-(4-Methyl)- piperazinylfullerene[60] - containing polysiloxane

Hendry Izaac Elim[a][*], Jianying Ouyang[b], Suat Hong Goh[b], and Wei Ji[a]
[a]Department of Physics, National University of Singapore, Singapore 117542
[b]Department of Chemistry, National University of Singapore, Singapore 117543
65-68742664, [*]scip1109@nus.edu.sg

*The nonlinear optical (NLO) properties of 1-(4-Methyl)- piperazinylfullerene[60] (MPF) incorporated with a polysiloxane copolymer denoted as PSI-46 have been studied using nanosecond laser pulses at 532-nm wavelength. The optical limiting performance of MPF itself is poorer than that of its parent $C_{60}$, while the contribution of PSI-46 attached to the MPF shows marginally improvement from the ratio of MPF/PSI-46(1:2) to MPF/PSI-46(1:6). However, the effect can be neglected to the MPF optical limiting responses. The photoluminescence emission of MPF-containing polymers was slightly improved in comparison with that of MPF. The possible sources for the slightly improvement in the NLO behaviour are discussed.*

Inspired by the first report of supramolecular fullerene introduced by Ermer in 1991 [1], many related researches have been increasing particularly about the formation of supramolecular buckminsterfullerene and fullerene chemistry, assemblies and arrays held together by weak intermolecular interactions [2-5]. Furthermore, the recent studies deal with supramolecular $C_{60}$-containing polymeric materials based on functionalized $C_{60}$ and polymers possessing suitable functional groups, which successfully overcome the incompatibility between pristine $C_{60}$ and polymers. The interactions between complementary functional groups of $C_{60}$ derivatives and polymers enable the $C_{60}$ derivatives to be well dispersed and adhere strongly to the polymer matrixes, leading to a significant improvement on the storage moduli of the materials [6,7]. In addition, optical limiting (OL) studies of multifunctional fullerenol incorporated with poly(styrene-co-4-vinylpyridine) were found to be poorer than that of its parent $C_{60}$ [8]. In contrast, the monofunctional 1,2-dihydro-1,2-methanofullerene [60]-61-carboxylic acid (FCA) –containing poly(styrene-co-4-vinylpyridiene) (PSVPy32) showed better OL behaviour than that of $C_{60}$ due to its high triplet-triplet state absorption [9]. However, the nonlinear optical (NLO) properties of a new supramolecular fullerene incorporated with polymers (PSI-46) have not been completely received serious attention yet. Therefore, our focus in the present work is to study the NLO performances of supramolecular $C_{60}$-containing polysiloxane based on carboxylated poly(dimethylsiloxane) and a multifunctional 1-(4-Methyl)- piperazinylfullerene[60] (MPF). It was envisaged that interaction between the carboxyl and amine groups would lead to supramolecular closslinks with MPF serving as physical crosslinking points. Poly(dimethylsiloxane) (PDMS) is a flexible polymer with low glass transition temperature, very low surface energy, low gas permeability, good thermal stability and biocompatibility [10,11]. It is desirable to combine the outstanding properties of PDMS and $C_{60}$. However, the multiple addends of MPF may distort the molecular symmetry of [60]fullerene, leading to significant changes of the NLO behaviors.

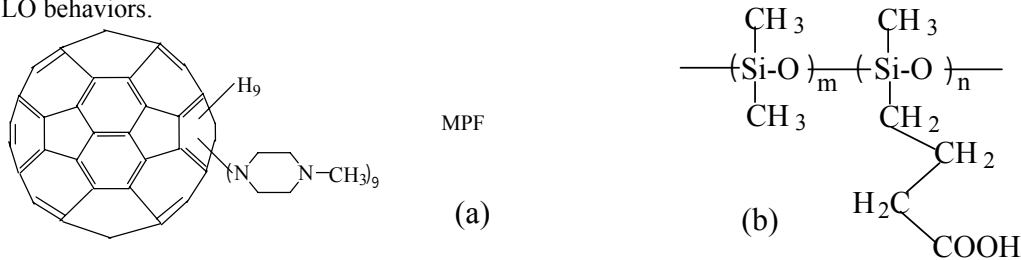

Figure 1. The structures of (a) MPF and (b) PSI-46.

Supramolecular 1-(4-Methyl)- piperazinylfullerene - containing polymers was prepared based on the following steps. [60]Fullerene ($C_{60}$) (99.9%) was obtained from Peking University, China. 1-Methylpiperazine (98%) was purchased from Sigma-Aldrich Company, USA. (3-Cyanopropyl)methyldichlorosilane and dichlorodimethylsilane were supplied by Fluka Chemika-Biochemika Company. Chlorobenzene (AR grade) was purchased from BDH, UK, and tetrahydrofuran (THF, AR grade) from Fisher Scientific, UK. 1-(4-Methyl)- piperazinylfullerene (MPF), a red and powder-like multifunctional $C_{60}$ derivative, was synthesized and characterized according to literature [12], which has an average stoichiometry of $C_{60}(HNC_4H_8NCH_3)_9$ with a structure as displayed in Fig. 1(a). The random copolymer of dimethylsiloxane and (3-carboxypropyl)methylsiloxane was synthesized and characterized according to the method reported by Li and Goh [13]. It contains 45.5 mol% of (3-

carboxypropyl)methylsiloxane unit as determined by $^1$H-NMR and has a number-average molecular weight 4,100 and polydispersity 1.08. This transparent and gel-like copolymer is denoted as PSI-46 (Fig. 1(b)). To prepare supramolecular MPF/PSI-46 composites, an appropriate amount of PSI-46 was added into the THF solution of MPF. The mixture was continuously stirred overnight. Three samples were prepared, which are MPF/PSI-46 (1:2), MPF/PSI-46 (1:4), MPF/PSI-46 (1:6), which are denoted as MPF (1:2), MPF (1:4), and MPF (1:6), respectively, where the ratios in the parentheses refer to the ratios of nitrogen atoms of MPF over the carboxylic groups of PSI-46.

The ultraviolet and visible (UV-vis) absorption sectra were recorded at the wavelength range 190-820 nm on a Hewlett Packard 8452A Diode Array spectrophotometer with a Hewlett Packard Vectra QS/165 computer system. The optical limiting measurements were conducted using linearly polarized nanosecond optical pulses from a Q-switched, frequency-doubled Nd:YAG laser (Spectra Physics DCR3) with pulse duration of 7 ns or an optical parametric oscillator (Spectra Physics MOPO 710) with pulse duration of 5 ns. The procedure of a detail OL measurement has been explained elsewhere [9]. While the photoluminescence (PL) measurement for each sample was managed using a luminescence spectrometer (LS 55, Perkin-Elmer Instrument U.K.) with an excitation wavelength of 480 nm.

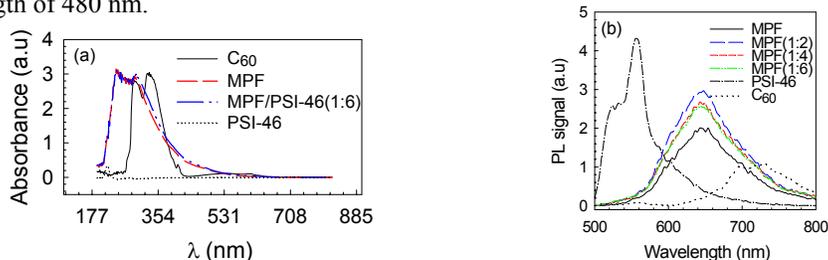

Figure 2. (**a**) UV-Vis spectra of $C_{60}$ in chlorobenzene and various samples of MPF/PSI-46 in tetrahydrofuran (THF). All the samples have the same linear transmittance of 75% at 532 nm through 1 mm-optical length. (**b**) Photoluminescence signals of 10-mm-thick PSI-46, $C_{60}$ in chlorobenzene and MPF, MPF(1:2), MPF(1:4), and MPF(1:6) in tetrahydrofuran (THF), respectively measured at the same excitation source of 480 nm.

Figure 2(a) shows the ground-state absorption spectra of PSI-46, MPF, and MPF/PSI-46 mixture solutions in THF. The absorption spectrum of $C_{60}$ in chlorobenzene is also shown for comparison. All the MPF samples have concentration of 9.05 x $10^{-3}$ M, and transmittance of ~75% at 532 nm, respectively. The molar absorptivity ($\varepsilon$) of MPF and MPF/PSI-46 is 140 M$^{-1}$cm$^{-1}$, which is near to 50 M$^{-1}$cm$^{-1}$ of the hexakis-adducts of methano-$C_{60}$ derivative [15], but much lower than that of 1080 M$^{-1}$cm$^{-1}$ of Fol [8]. PSI-46 basically shows no absorption, and MPF exhibits two absorption peaks at 246 and 300 nm, respectively, locating at lower wavelengths as compared to those of $C_{60}$, which may be ascribed to the weakening π-conjugation caused by multi-addends of MPF. Moreover, the spectrum of MPF is structureless in the visible region, which is different from that of $C_{60}$. The addition of PSI-46 has negligible effects on the absorption spectrum and the linear transmittance at 532 of MPF. Figure 2(b) displays that there is a significant shift of the photoluminescence (PL) peak of MPF and MPF/PSI-46 in comparison with that of $C_{60}$. This may be caused by the strong chemical bonds due to the supramolecular which distorts the molecular symmetry of fullerenes. Moreover, there is an increase in the PL intensity of MPF incorporated with PSI-46, in comparison to its parent MPF. This is due to a significant disturbance of π-electron system of supramolecular MPF upon derivatization, consistent with our UV-vis data. Slightly enhanced photoluminescence of low amount of PSI-46 for these supramolecular-containing polymers can be attributed to a decrease in the rate of intersystem crossing ($k_{isc}$) between the first excited singlet and triplet states induced by the reduction in symmetry as a result of the partially broken π system. Derivatisation which gives rise to a decrease in the symmetry of the fullerene converts some vibronic forbidden states to the allowed states, thus increasing their population in the first excited singlet state and transition probabilities which results in decreasing of the rate constant value of the intersystem crossing $k_{isc}$. Thus, the increase in fluorescence is explained by a decrease in the probability of ($S_1 \rightarrow T_1$) radiationless transitions. However, by adding more PSI-46 into the MPF, the PL becomes marginally diminished. The higher the polymer content is, the slightly lower the PL intensity is. This implies a reduction in the emission from the first excited singlet state to the ground state. Consequently, it increases the probability of transferring electrons to the lowest excited triplet state by inter-system crossing. In addition, it should be noted that the PL emission of $C_{60}$ in chlorobenzene still exhibits very weak luminescence at room temperature due to high molecular symmetry of $C_{60}$ in comparison with MPF and MPF/PSI-46.

As shown in Figure 3(a), the optical limiting response of MPF is weaker than $C_{60}$, which may be due to the disturbance of π-conjugation of $C_{60}$ cage by multi-addends of MPF. And the optical limiting responses of MPF/PSI-46 mixtures are marginally weaker than that of MPF. However, by adding more PSI-46 into MPF, for instance from MPF/PSI-46(1:2) to MPF/PSI-46(1:6), there is a little improvement of the OL of MPF-containing polymer. The limiting behavior is attributed predominantly to the absorption cross section of the lowest excited triplet state. Furthermore, the triplet-triplet absorption in supramolecular MPF/PSI-46 has been gained by attaching more content of polysiloxane. This indicates that the high polysiloxane concentration may give marginally effects both on the excited triplet state properties and triplet-state lifetime of the MPF/PSI-46. Moreover, the limiting threshold of MPF (8 J/cm$^2$) extracted from Fig. 3(a) is slightly lower than that of MPF incorporated with PSI-46 (> 8 J/cm$^2$). To check photostability of the samples, we have measured and compared the absorption spectra before and after laser irradiation. The obtained results indicated that there is no difference in the spectra for all the samples. Therefore, all the samples have a good photostability.

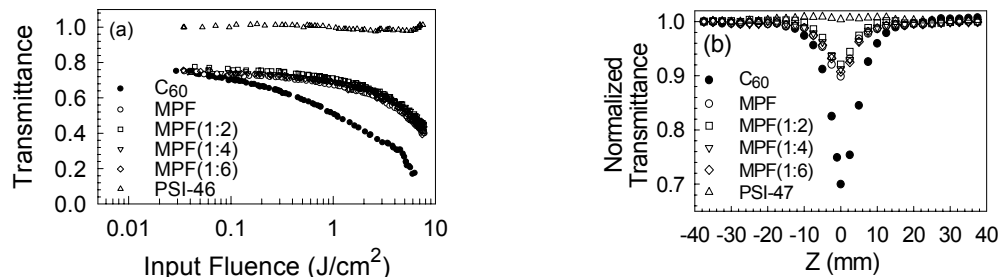

Figure 3. (**a**) Nonlinear transmission of $C_{60}$-chlorobenzene (filled circles), and MPF (open circles), MPF/PSI-46(1:2) (open squares) denoted as MPF(1:2), MPF(1:4) (open inverted triangles), MPF(1:6) (open diamonds), and PSI-46 (open triangles) in tetrahydrofuran (THF), respectively. The measurements are conducted with 7-ns laser pulses at λ = 532 nm. All the samples have the same linear transmission of $T$ = 75 %, and the optical path of 1 mm. (**b**) Open aperture Z-scans conducted at the input irradiance, $I$ = 170 MW/cm$^2$ and beam waist of 45 μm on 1-mm-thick solutions of $C_{60}$-chlorobenzene (filled circles), and MPF (open circles), MPF(1:2) (open squares), MPF(1:4) (open inverted triangles), MPF(1:6) (open diamonds), and PSI-46 (open triangles) in tetrahydrofuran (THF), respectively.

To assess the potential application of supramolecular 1-(4-Methyl)- piperazinylfullerene and MPF-containing PSI-46, nonlinear optical (NLO) measurements were carried out. Figure 3(b) shows the open aperture Z-scans [14] of $C_{60}$ in chlorobenzene, and MPF and MPF/PSI-46 in THF, respectively. They have been measured by using 5-ns laser pulses of λ = 532 nm at the input irradiance of 170 MW/cm$^2$. One can obtain that there is a significant difference of the signal between $C_{60}$ and MPF, which indicates the possibility of the disturbance of π–electronic system in $C_{60}$ cage due to multi-addends of MPF. This observation is consistent with that in OL measurements. However, the open aperture Z-scans of MPF/PSI-46 with different concentration of PSI-46 have a marginally alteration in the nonlinear absorption, in agreement with the limiting measurements in Fig. 3(a). The strongest Z-scan signal among them is observed in the MPF(1:6) sample. It suggests that the MPF with high concentration of PSI-46 has only had a small NLO effect and has been still poorer than that of MPF. To ensure there is no an existence of small optical nonlinearity contributed by PSI-46 in THF, we have carried out the Z-scan and have observed no nonlinear signal, as shown in Fig. 3(b). A closed relationship between optical limiting and nonlinear optical properties of the supramolecular MPF-containing polymers indicates that photo-excited electrons and their dynamic process play important role.

In conclusion, the optical limiting, Z-scan and PL measurements have been carried out to study the nonlinear optical properties of supramolecular 1-(4-Methyl)- piperazinylfullerene[60] - containing polysiloxane. The MPF and MPF incorporated with polymer possess poorer nonlinear optical properties than that of $C_{60}$, which are mainly caused by the disturbance of π-electron system in $C_{60}$ cage due to multifunctionalization.

**References**
[1] O. Ermer, *Helv. Chim. Acta* **74**, 1339 (1991).
[2] H. W. Kroto, J. R. Heath, S. C. O'Brien, R. F. Curl and R. E. Smalley, *Nature* **318**, 162 (1985).
[3] W. Krätschmer, L. D. Lamb, K. Fostiropoulos and D. R. Huffman, *Nature* **347**, 354 (1990).
[4] F. Diederich and C. Thilgen, *Science* **271**, 317 (1996).
[5] F. Diederich and M. Gómez-López, *Chem. Soc. Rev.* **28**, 263–277 (1999).
[6] J.Y. Ouyang, S.H. Goh and Y. Li, *Chem Phys Letts.* **347**, 344-348 (2001).
[7] J.Y. Ouyang and S.H. Goh, *Fullerenes, Natotubes, and Carbon Natonubes* **10**, 183-196 (2002).
[8] J.Y. Ouyang, S.H. Goh, H.I. Elim, G.C. Meng and W. Ji, *Chem Phys Letts.* **366**, 224-230 (2002).
[9] H.I. Elim, J.Y. Ouyang, J. He, S.H. Goh, S.H. Tang and W. Ji, *Chem Phys Letts.* **369**, 281 (2003).
[10] S.J. Clarson and J.A. Semlyen, *Siloxane polymers.* NJ: Prentice Hall, 216-244 (1993).
[11] R.G. Jones, W. Ando and J. Chojnowski, *Silicon-containing polymers.* The Netherlands: Kluwer Academic Publishers, 185-243 (2000).
[12] S.H. Goh, S.Y. Lee, Z.H. Lu and C.H.A. Huan, *Macromol. Chem. Phys.* **201**, 1037-1047 (2000).
[13] X. Li, S.H. Goh, Y.H. Lai, and A.T.S. Wee, *Polymer* **41**, 6563-6571 (2000).
[14] M. Sheik-Bahae, A. A. Said, T. Wei, D. J. Hagan and E. W. Van Stryland, *IEEE J. Quantum Electron* **26**, 760 (1990).
[15] Y.P. Sun and J.E. Riggs, *J. Chem. Phys.* **112**, 4221 (2000).